\documentclass{article}
\usepackage[utf8]{inputenc}
\usepackage{graphics,amsmath,amssymb,xcolor,pstool,amsthm,mathtools,soul,float}
\newtheorem*{remark}{Theorem}
\usepackage{accents}
\newcommand\thickbar[1]{\accentset{\rule{.5em}{.7pt}}{#1}}
\usepackage[colorlinks = true,
            linkcolor = blue!70!black,
            urlcolor  = red!70!black,
            citecolor = green!55!black,
            anchorcolor = blue!70!black]{hyperref}

\DeclareMathAlphabet\mathbfcal{OMS}{cmsy}{b}{n}
\DeclareSymbolFont{bbold}{U}{bbold}{m}{n}
\DeclareSymbolFontAlphabet{\mathbbold}{bbold}

\def\Nvac{\mathcal{N}_\text{vac}}
\def\Nred{\mathcal{N}_P}
\def\di{\mathrm{d}}
\def\qalpha{\boldsymbol{q}^{(\alpha)}}
\def\qalphaT{\boldsymbol{q}^{(\alpha) \top}}
\def\qalphaprime{\boldsymbol{q}^{(\alpha')}}
\def\Tvec{\boldsymbol{T}}
\def\tauvec{\boldsymbol{\tau}}
\def\thetavec{\boldsymbol{\theta}}
\def\thetatildevec{\widetilde{\boldsymbol{\theta}}}
\def\Qtilde{\widetilde{\mathbfcal{Q}}}
\def\Qbar{\thickbar{\mathbfcal{Q}}}
\def\Lambdatilde{\widetilde{\Lambda}}
\def\Lambdabar{\thickbar{\Lambda}}
\def\deltavec{\boldsymbol{\delta}}
\def\deltatilde{\widetilde{\boldsymbol{\delta}}}
\def\deltabar{\thickbar{\boldsymbol{\delta}}}
\def\alphavec{\boldsymbol{\alpha}}
\def\qbar{\thickbar{\boldsymbol{q}}}

\begin{document}
\begin{center}
{\Large \bf Axion minima in string theory} \\ [0.7cm]

{Naomi Gendler$^{\,\rm a, \rm b}$, Oliver Janssen$^{\,\rm c, \rm d, \rm e}$, Matthew Kleban$^{\,\rm f}$}, Joan La Madrid$^{\,\rm f}$ and Viraf M. Mehta$^{\,\rm g}$
\\ [0.5cm]

\small{
\textit{$^{\rm a}$
Jefferson Physical Laboratory, Harvard University, Cambridge, MA  02138, USA}}
\vspace{.2cm}

\small{
\textit{$^{\rm b}$
Department of Physics, Cornell University, Ithaca, NY  14850, USA}}
\vspace{.2cm}

\small{
\textit{$^{\rm c}$
International Centre for Theoretical Physics, Strada Costiera 11, 34151, Trieste, Italy}}
\vspace{.2cm}

\small{
\textit{$^{\rm d}$
Institute for Fundamental Physics of the Universe, Via Beirut 2, 34014, Trieste, Italy}}
\vspace{.2cm}

\small{
\textit{$^{\rm e}$
Laboratory for Theoretical Fundamental Physics, EPFL, BSP 530 Rte de la Sorge, CH-1015, Lausanne, Switzerland}}
\vspace{-.2cm}

\small{
\textit{$^{\rm f}$
Center for Cosmology and Particle Physics, New York University, 726 Broadway, New York, NY
10003, USA}}
\vspace{.2cm}

\small{
\textit{$^{\rm g}$
Institut f\"ur Astrophysik, Georg-August Universit\"at, Friedrich-Hund-Platz 1, G\"ottingen, Germany}}

\end{center}

\vspace{0.5cm}

{\sc Abstract:} We study the  landscape of axion theories in compactifications of type IIB string theory on orientifolds of Calabi-Yau threefolds. In a  sample of approximately 400,000 geometries we find that in the regime of perturbative control there are only a handful of distinct axion minima per geometry, despite there being infinitely many instanton contributions to the potential with unbounded charges. The ensemble we consider has numbers of axion fields ranging from 1 to 491, but the median number of distinct minima is 1, the mean number is 1.9 and the largest is 54. These small numbers of minima occur because the leading axion charge matrix is quite sparse, while the subleading corrections are increasingly exponentially suppressed as the charges increase.  On their own, such potentials are nowhere near rich enough to be of interest anthropically.  This is in stark contrast to potentials for which the charge matrix is less sparse or the hierarchies between the instanton contributions are less steep, where one can find $\mathcal{O}(10^{500})$ minima for $\mathcal{O}(500)$ axions.  To generate a sufficiently large landscape from string compactifications our results indicate that one would need to rely on varying flux or topology, or to develop tools that allow one to go beyond the regime we can control with current techniques.

\newpage
\tableofcontents

\section{Introduction}
Since the work of Kaluza and Klein it has been known that compactifying extra dimensions gives rise to fields that transform as scalars or pseudoscalars in the four non-compact spacetime dimensions. String theory compactifications in particular often have hundreds of axions \cite{Peccei:1977hh,PhysRevLett.40.279,PhysRevLett.40.223} -- pseudoscalar fields with potentials that are periodic in all directions of the field space \cite{Wen:1985jz,Svrcek:2006yi}. Within the multi-dimensional fundamental domain defined by these periods the potential can be very complex: when the parameters of the theory are chosen randomly, the potential can have $\Nvac \sim 10^\text{hundreds}$ of unique local minima (along with a rich set of other features) \cite{Bachlechner:2017zpb, Bachlechner:2017hsj} and a correspondingly dense ``discretuum"  of values for the vacuum energy with spacing $1/\Nvac$ \cite{Bachlechner:2015gwa}.   Such complex energy landscapes are extremely interesting. They can solve the cosmological constant problem anthropically \cite{Weinberg:1987dv, Bousso:2000xa} and account for many of the other macroscopic features of our universe, including the Big Bang, dark matter, and early-universe inflation \cite{Bachlechner:2019vcb}.

 However not all axion theories -- even those with hundreds of fields -- give rise to rich landscapes with large numbers of minima.  In this work we investigate a large class of Calabi-Yau (CY) compactifications of string theory for which data on axion potentials is available \cite{Demirtas:2018akl}. As we will see, even in compactifications with $\sim 500$ axions, in almost all the parameter space there are relatively few distinct minima -- far too few for an anthropic approach to solving the cosmological constant problem to succeed based on the axion potential for an individual compactification.  Instead, our results indicate that any such approach  must rely on varying topology or flux \cite{Denef:2004ze}.
 
\paragraph{An example from field theory} To gain some intuition regarding axion potentials, consider field theory in 5D with an Abelian gauge field $A$, some charged matter, and the extra dimension compactified on a circle of circumference $L$. Due to gauge invariance, the fifth component $A_5$ of the gauge field cannot have a local potential. However, the charged matter will generate a non-local potential for the Wilson loop
\begin{equation}
\label{eq:WL}
e^{i \theta} = \exp \left( i \oint \di x^5 \, A_5 \right) \,.
\end{equation}
At energies below the Kaluza-Klein scale the axion $\theta$ gets a potential $V$ given at one-loop by (see \cite{Arkani-Hamed:2003xts} and references therein)
\begin{equation}
\label{gaugeax}
V(\theta) =  \frac{C}{L^4} 
\sum_{q=1}^{\infty} e^{- q m L}
\left( \frac{1}{q^5} + \frac{mL}{q^4} + \frac{(mL)^2}{3q^3} \right) \cos(q g \theta) \,,
\end{equation}
where the dimensionless constant $C$ depends on the characteristics of the matter field with charge $g$ and mass $m$. The potential is a periodic Fourier series for $\theta$ with coefficients that are exponentially suppressed in  $q m L$, proportional to the frequency $q$ (or ``axion charge"). One can imagine the worldline of the charged particle wrapping $q$ times around the  cycle; the suppression comes from the worldline action, mass times length. In string theory the axion potential has a very similar overall structure, consisting of periodic terms in the potential with frequency $\propto q$ exponentially suppressed by the action for a brane wrapping $q$  times around the cycle, brane tension times volume. There are many topologically distinct cycles, giving rise to many axion fields and many different types of contributions to the potential.  When the volume of the compact manifold is large in string units and $g_s \ll 1$ the potential is exponentially dominated by its first few terms and one can generally ignore the rest of the sum.

A glance at \eqref{gaugeax} shows that at large $mL$ where all but the leading $q=1$ term can be neglected, there is a single distinct minimum\footnote{There are infinitely many minima on the real line, but there is a single one per period $2 \pi/g$.} located at $\theta = 0$ mod $2 \pi/g$.  A richer structure of minima could have  occurred in two ways.  One is if the largest term in the potential had higher frequency than some subdominant terms.  For instance, $\cos{10 \, \theta} + \varepsilon \cos \theta$ has 10 unique minima for small $\varepsilon \neq 0$ because the second term lifts a degeneracy and expands the fundamental domain from $2 \pi/10$ to $2\pi$ \cite{Bachlechner:2015gwa}. The other way is to have several terms with different frequencies and comparable coefficients that interact to form a more complex energy landscape \cite{Bachlechner:2017hsj}. Neither of these effects occurs in \eqref{gaugeax}.\footnote{In fact, the $q^{-5}$ suppression in \eqref{gaugeax} is strong enough that there is only one minimum even for small $m L$, but a lesser suppression would have admitted additional minima.} As we will see, neither occurs in any significant way in the class of string theory compactifications we will study either. This is the fundamental reason we do not find a rich axion potential with many distinct minima.

\section{Axion potentials in string theory} \label{stringsec}
We are interested in studying vacua in the landscape of axion potentials generated in string theory. Understanding this landscape in full generality is currently an intractable task: it requires understanding the full set of geometric string theory solutions, as well as transitions between vacua arising from different geometries. In this work, we make progress by focusing on a subset of these solutions: we study vacua in compactifications of type IIB string theory on O3/O7 orientifolds of CY threefolds.\footnote{We do not explicitly perform orientifolds of CY threefolds, but projecting out instantons would generally reduce the number of minima.}

The matter content of the low energy effective theories arising from such compactifications includes $h^{1,1} \equiv N$ axions, as well as their scalar counterparts, called saxions. At the $\mathcal{N}=2$ level, in the four-dimensional effective field theory, these particles are massless, and in particular the saxions will give rise to new forces that are stringently constrained by ``fifth force" experiments. In order to obtain solutions of string theory that are potentially compatible with observations of the real world, these light fields must be given a mass. In type IIB string theory, saxions can obtain a potential from non-perturbative effects -- Euclidean D3-branes or gaugino condensation  wrapped on compact cycles in the CY -- or perturbative corrections to the K\"ahler potential. 
   In our work, we are assuming that there are perturbative effects that stabilize the saxions at any point in moduli space we consider.

The same non-perturbative ingredients that contribute to the potential for the saxions give a potential to the axions. The theory is determined in terms of the K\"ahler potential $\mathcal{K}$, given below at tree level, and a (non-perturbative) superpotential $W$:
\begin{align}
    \mathcal{K} &= - 2\log \mathcal{V} \,, \label{KahlerpotentialEq} \\
    W &= W_0 + \sum_{\alpha} A_\alpha \exp(-2\pi \qalpha \cdot \Tvec) \,, \label{superpotentialEq}
\end{align}
where the sum is taken over all instantons, of which there are presumably infinitely many (but see \cite{Alexandrov:2022m}), $\mathcal{V}$ is the overall volume of the CY, $W_0$ is the perturbative flux Gukov-Vafa-Witten superpotential \cite{Gukov:1999ya}, $A_\alpha$ are one-loop fluctuation determinants, $\qalpha = q^{(\alpha)}_i$ are the instanton charges and $\Tvec$ are the K\"ahler moduli\footnote{Matrices and vectors (by definition, columns) will be written in boldface and indices are always contracted with $\mathbbold{1}$, e.g.~$\boldsymbol{q}^{(\alpha) \top} \Tvec = \qalpha \cdot \Tvec = \delta^{ij} q^{(\alpha)}_i T_j = q^{(\alpha)}_i T^i$. We measure volumes in Einstein frame, in units where $\ell_s=1$.}, whose real and imaginary parts are the saxions and axions, respectively:
\begin{equation}
    T^i = \tau^i + i \theta^i.
\end{equation}
Indices $i,j$ run over the K\"ahler moduli from 1 to $N = h^{1,1}$. Importantly, the saxions $\tau^i$ are the volumes of holomorphic four-cycles in the internal manifold, i.e. cycles that are calibrated by the K\"ahler form $J$:
\begin{equation}
    \tau^i = \frac{1}{2} \int_{D_i} J \wedge J \,,
\end{equation}
where $D_i$ is a divisor (four-cycle) in the CY. $J$ takes values in the K\"ahler cone of the CY, defined as the region where all calibrated cycles have non-negative volumes. The axions $\theta^i$ are obtained by integrating the ten-dimensional Ramond-Ramond four-form gauge field $C_4$ over this same divisor:
\begin{align}
    \theta^i = \int_{D_i} C_4 \,.
\end{align}
The shift symmetry of the axion is thus inherited from the gauge symmetry of the higher-dimensional gauge field. 

Given $\mathcal{K}$ and $W$ as in \eqref{KahlerpotentialEq} and \eqref{superpotentialEq}, we can compute the  F-term potential from supergravity:
\begin{equation}
    V = e^\mathcal{K} \left( K^{a \overline{b}} D_a W \overline{D_b W} - 3|W|^2 \right) \,,
\end{equation}
where the covariant derivative is $D_a \equiv \partial_a + \partial_a \mathcal{K}$ and $K^{a \overline{b}}$ is the inverse K\"ahler metric associated to $\mathcal{K}$. Plugging in \eqref{KahlerpotentialEq} and \eqref{superpotentialEq}, we arrive at the following form for the axion potential \cite{Demirtas:2018akl}:\footnote{Following \cite{Demirtas:2018akl} we will make the choice $|W_0| = |A_\alpha| = \mathcal{O}(1)$.}
\begin{align} \label{Vfull}
V(\thetavec) &= -\frac{8 \pi}{\mathcal{V}^2} \left[ \sum_{\alpha} (\qalpha \cdot \tauvec) \, e^{-2 \pi \qalpha \cdot \tauvec} \cos \left(2 \pi \qalpha \cdot \thetavec + \delta_\alpha \right) \right. \nonumber \\ &+ \frac{1}{2}
\left.\sum_{\alpha \neq \alpha'}\left( \pi \, \qalphaT \boldsymbol{K}^{-1} \qalphaprime + \left( \qalpha + \qalphaprime \right) \cdot \tauvec \right) e^{-2 \pi \left( \qalpha + \qalphaprime \right) \cdot \tauvec} \cos \left( 2 \pi \left( \qalpha - \qalphaprime \right) \cdot \thetavec + \delta_{\alpha, \alpha'} \right) \right] \,,
\end{align}
where the phases $\delta_\alpha, \delta_{\alpha, \alpha'}$ are determined by the phases of $W_0$ and $A_\alpha$.\footnote{The Lagrangian for $\thetavec$ includes a non-canonical kinetic term involving $\boldsymbol{K}$, but, as we are only interested in the vacuum structure of the theory here (the minima of $V$) the normalization of $\thetavec$ is unimportant.} 

To determine the set of instanton charges $\{ \qalpha \}$, one needs to know which divisors in the CY can host instantons that contribute to $\mathcal{K}$ and $W$. We will focus here on non-perturbative contributions to $W$, leaving a discussion of the corrections to $\mathcal{K}$ to Appendix \ref{Kahlerappendix}. $W$ is a holomorphic object, therefore Euclidean D3-branes can only give contributions to it if they are BPS, i.e., if they are wrapped on holomorphic four-cycles.\footnote{We are ignoring the possibility of gaugino condensation contributing to the potential.  We do not expect this to affect our qualitative conclusions, as to account for it it would be reasonable to rescale the divisor volumes by a constant factor.} Determining the set of charges is therefore reduced to finding the set of calibrated divisors. Determining whether any particular instanton appears with a non-vanishing one-loop determinant $A_\alpha$ is more subtle, however, requiring the counting of the fermion zero modes of that instanton. We will not attempt this task here -- instead, we will assume that every divisor that can host a superpotential instanton based only on holomorphicity indeed induces a contribution.\footnote{On general grounds we expect that including too many contributions to $V$ will \textit{overcount} the amount of minima, although this is not a theorem.}  

The charges of divisors that are calibrated in a CY threefold are given by the points in an integral cone known as the effective cone, $\mathcal{E}$. The charges in $\mathcal{E}$ are the charges that appear in the effective potential \eqref{Vfull} (modulo fermion zero modes). In general, determining the exact effective cone for a given geometry is a difficult task (see \cite{Gendler:2022qof,Gendler:2022ztv} for progress in this direction). For CY threefolds constructed as hypersurfaces in toric varieties (to which we shall restrict), however, a particular set of effective divisors is guaranteed: these are known as the prime toric effective divisors. For ``favorable" CYs\footnote{Those for which all prime toric divisors on the ambient toric variety which intersect the CY restrict to irreducible divisors on the CY.}, there are exactly $h^{1,1} + 4 = N + 4$ prime toric divisors. These divisors are inherited from the ambient toric variety. In addition to these, there can be effective divisors that are \textit{not} inherited from the ambient variety -- these are known as ``autochthonous" divisors \cite{Demirtas:2018akl}. There is no bound on how many autochthonous divisors a particular geometry might have: in fact, see \cite{Gendler:2022qof,Donagi:1996yf} for examples with infinitely many. A rigorous determination of the vacuum structure requires knowledge of the entire set of autochthonous divisors. However, autochthonous divisors can generally be neglected as previous work has found that the the corresponding cycle volumes are relatively large \cite{Demirtas:2021nlu}.

In the course of this work we discovered a class of divisors that appears to have gone unnoticed in previous research on this topic.  The intersection of the cone generated by the prime toric divisors (PTDs) with the integer lattice in general contains points that cannot be represented by non-negative integer linear combinations of the PTDs.  Instead, these points correspond to fractional linear combinations of the PTDs.  These points can be generated by a larger basis of integer vectors (the unique, minimal such set is known as the Hilbert basis).  It remains unclear if these points are effective, that is, if they correspond to calibrated cycles and how they contribute to the axion potential.  While computing the Hilbert basis is computationally intractable at large $N$, we developed an algorithm that allowed us to investigate the effects of  including these points in the axion potential at all $N$.  After a preliminary scan over a sample of geometries,  it appears including them in the potential does not affect our conclusions qualitatively, and thus we construct axion potentials using only the PTDs.  We leave further investigation of this subtlety to future work.

After making these simplifying assumptions, we are left with an approximation to $\mathcal{E}$ that is generated by $N+4$ effective divisors. To write the effective potential explicitly we  pick a basis of these divisors in which they take following form:
\begin{align} \label{primetoric}
    \begin{pmatrix} \mathbbold{1}_N \\
    \boldsymbol{q}^{(1) \top} \\
    \boldsymbol{q}^{(2) \top} \\
    \boldsymbol{q}^{(3) \top} \\
    \boldsymbol{q}^{(4) \top}
\end{pmatrix} \,,
\end{align}
where the $\boldsymbol{q}^{(1,\cdots,4)}$ are  integer-valued.  Since these vectors generate a cone, the infinitely many charge vectors in the first line of \eqref{Vfull} include all possible  linear combinations of these $N+4$ charges with \emph{non-negative} integer coefficients.  These  will be exponentially suppressed in the potential compared to the $N+4$ charges in \eqref{primetoric} so long as the divisor volumes are not small in string units.

\subsection{Axion fundamentals}\label{axfund}
The potential \eqref{Vfull} is invariant under $N$ linearly independent symmetries; discrete shifts in the axion field space. These shifts define a periodic lattice in $N$ dimensions that can be characterized by a fundamental domain which has the smallest possible volume and tiles the full lattice.\footnote{See \cite{Bachlechner:2017hsj} for a detailed discussion. The fundamental domain is not unique -- lattice bases related via multiplication by a unimodular matrix (integer-valued with determinant one) are equivalent.} One finds the distinct minima of the potential by locating all the minima inside one fundamental domain. Since the identity matrix is contained among the charges $\{ \qalpha \}$ (recall Eq.~\eqref{primetoric}), a candidate fundamental domain of the axion potential \eqref{Vfull} is the unit $N$-cube, $\mathcal{F}(V) = [0,1)^N$. If all other charges contain integers only, this is indeed a fundamental domain. (This is the generalization to $N > 1$ of the simple example considered in the introduction.) If there are fractions  the fundamental domain can have larger than unit volume.

In \S\ref{numericalmin} we will explain how to truncate the potential down to $P \geq N$ terms that constitute a sufficiently accurate approximation to the full potential, so that
\begin{align} \label{VP}
V(\thetavec) \approx V_P(\thetavec) = \sum_{I=1}^P \Lambda_I^4 \, \cos \left(2 \pi (\mathbfcal{Q} \thetavec)^I + \delta_I \right) \,.
\end{align}
Here $\mathbfcal{Q}$ will be a full rank $P \times N$ matrix containing $P$ charge vectors $\boldsymbol{q}^{(I)}$ on its rows, the $\Lambda_I^4, \delta_I$ are the corresponding coefficients and phases of the cosines derived from Eq.~\eqref{Vfull}, and we have ignored an additive constant. In the regime where where the volumes $\tau^i$ in \eqref{Vfull} are collectively scaled to infinity and perturbative control becomes arbitrarily precise (see \S\ref{perturbativesec}), $P = N$. Away from this limit but still in the regime of control one typically has $P \lesssim N+4$, where most of the $P$ rows in $\mathbfcal{Q}$ must be chosen from the those in Eq.~\eqref{primetoric}, so that $\mathbfcal{Q}$ is a sparse matrix.

The nature of the truncation \eqref{VP} guarantees that the total number of distinct minima of $V$ is equal to an integer multiple of the number of distinct minima of $V_P$. This follows from the fact that each distinct minimum of $V_P$ is a good approximation to a distinct minimum of $V$. However, in truncating $V$ to $V_P$ we may have decreased the volume of the fundamental domain, so that identical minima of $V_P$ could correspond to distinct minima of $V$ (where the degeneracy is lifted by the subleading terms we have neglected when truncating, just as in the example we discussed in the introduction). The ratio of the number of distinct minima of $V$ to that of $V_P$ is equal to the ratio of the volumes of their respective fundamental domains, $\text{vol} \, \mathcal{F}(V) / \text{vol} \, \mathcal{F}(V_P)$.\footnote{We prove in Appendix \ref{integervacuaappendix} that this ratio is always an integer. Notice that the volumes separately are not coordinate-independent but that their ratio is.} Thus, denoting by $\Nvac$ the number of distinct minima of $V$ and by $\Nred$ the number of distinct minima of $V_P$,
\begin{equation} \label{Nvaceq}
    \Nvac = \frac{\text{vol} \, \mathcal{F}(V)}{\text{vol} \, \mathcal{F}(V_P)} \Nred \,.
\end{equation}
More generally we expect the same proportionality factor to relate the amount of critical points of $V$ to those of $V_P$.

To proceed let us begin with the simplest case where $P=N$, so that the truncated potential \eqref{VP} consists of $N$ cosines with arguments that are linear combinations of the $N$ axion fields, viz.
\begin{align} \label{Vsimplest}
V_P(\thetavec) &= \sum_{I=1}^N \Lambda_I^4 \, \cos \left(2 \pi (\mathbfcal{Q} \thetavec)^I + \delta_I \right) \,.
\end{align}
In this case -- which describes the large volume limit --  we can easily determine how many unique minima $\Nvac$ the full potential has.  Consider the coordinate transformation\footnote{Recall $\mathbfcal{Q}$ has full rank (so here where $P = N$, is invertible), which will ensured by construction in \S\ref{numericalmin}.}
\begin{equation}
    2 \pi \mathbfcal{Q} \thetavec + \deltavec = 2 \pi \thetatildevec \,.
\end{equation}
In these coordinates the potential is sum-separable,
\begin{align} \label{Vsimplest2}
V_P(\thetatildevec) &= \sum_{I=1}^N \Lambda_I^4 \, \cos \left(2 \pi \widetilde{\theta}^I \right) \,,
\end{align}
and it becomes manifest there is only a single distinct minimum: $\Nred = 1$. It remains to determine the ratio of volumes in Eq.~\eqref{Nvaceq} to determine $\Nvac$. For simplicity let us specify to cases where $\mathcal{F}(V) = [0,1)^N$ (in the original $\thetavec$-coordinates), so $\text{vol} \, \mathcal{F}(V) = 1$. To find $\mathcal{F}(V_P)$ and its volume, we must find those integer $N$-vectors $\boldsymbol{n}$ for which there exists a $\boldsymbol{\theta}$ with $\mathbfcal{Q} \thetavec = \boldsymbol{n}$. Clearly, all $\boldsymbol{n} \in \mathbb{Z}^N$ satisfy this condition: $\boldsymbol{\theta} = \mathbfcal{Q}^{-1} \boldsymbol{n}$. A basis for $\mathbb{Z}^N$ is simply $\mathbbold{1}_N$, and so, a basis in the $\boldsymbol{\theta}$-coordinates consists of the columns $\boldsymbol{q}^{(i)}_{-1}$ of $\mathbfcal{Q}^{-1}$. A fundamental domain of the truncated potential is then
\begin{equation}
    \mathcal{F}(V_P) = \left\{ \sum_{i=1}^N p_i \boldsymbol{q}^{(i)}_{-1} ~ \big| ~ p_i \in [0,1) \right\}
\end{equation}
with volume \cite{latticeintro}
\begin{equation} \label{volFV}
    \text{vol} \, \mathcal{F}(V_P) = \left| \det \mathbfcal{Q}^{-1} \right| = 1 / \left| \det \mathbfcal{Q} \right| \,.
\end{equation}
Therefore in the bulk of our ensemble we have the exact result
\begin{equation} \label{exactNvac}
    \Nvac = \left| \det \mathbfcal{Q} \right| \,.
\end{equation}

When $P > N$ the exact amount of minima must be determined numerically; in \S\ref{criticalsec} we describe our procedure to locate critical points of \eqref{VP}, specializing to its distinct set of minima in \S\ref{minimasec}. This determines $\Nred$ in \eqref{Nvaceq}. Determining $\mathcal{F}(V_P)$ and its volume now requires finding a basis for the lattice of intersections between the hyperplane defined by $\{ \mathbfcal{Q} \thetavec \}$ and the integer $P$-lattice $\mathbb{Z}^P$. This can be done efficiently using the Smith normal decomposition of $\mathbfcal{Q}$, as we describe in Appendix \ref{Smithappendix}. None of these factors are substantial, however, since $\mathbfcal{Q}$ is filled only sparsely with $\mathcal{O}(1)$ entries. The paucity of minima thus persists for $P > N$.

\subsection{Computational control} \label{perturbativesec}
In writing down the axion potential \eqref{Vfull} that results from compactifying type IIB string theory on a Calabi-Yau threefold (orientifold), we need to ensure that the various expansion parameters are in regimes such that we can trust our evaluation of the effective theory.

The first consideration is that the $\alpha'$ expansion must be under control. As a proxy for this, we ensure that all two-cycles and four-cycles in the geometry have volumes bigger than a threshold that we take to be $1$ in string units. In other words, for each complex curve $\mathcal{C}$ and divisor $\mathcal{D}$, we guarantee that
\begin{align}
    \text{vol} \, \mathcal{C} \geq 1
    \label{eq:vol1}
\end{align}
and
\begin{align}
    \text{vol} \, \mathcal{D} \geq 1
\end{align}
in string units.

Parametric satisfaction of this condition is trivial to arrange: given an arbitrary K\"ahler form $J$ in the interior of the K\"ahler cone, we can simply scale $J \rightarrow \lambda J$ with $\lambda \rightarrow \infty$. Larger $\lambda$ corresponds to steeper hierarchies and therefore fewer relevant terms in the potential, which in general leads to simpler potentials with fewer distinct minima. Therefore we are interested in finding the point in moduli space that is both (i) in the regime of computational control and (ii) generally supports the largest number of minima.

The subregion of the K\"ahler cone where \eqref{eq:vol1} is satisfied is known as the ``stretched K\"ahler cone" \cite{Demirtas:2018akl}, constructed by translating the walls of the K\"ahler cone in by a distance $1$. The ``tip of the stretched K\"ahler cone" is then the point in the stretched K\"ahler cone that lies closest to the origin. This point can be found for any given geometry using an optimizer implemented in \texttt{CYTools} \cite{Demirtas:2022hqf}. In every geometry we consider, our starting point will be the tip of the stretched K\"ahler cone, so that we ensure convergence of the $\alpha'$ expansion.\footnote{Note that in general, imposing that \textit{every} curve has large volume is too restrictive: curves corresponding to flops can be smaller than the imposed threshold \cite{Demirtas:2021nlu, Gendler:2022ztv}. Therefore, it would be reasonable to uniformly sample the moduli space, and then do the smallest dilatation $J \rightarrow \lambda J$  possible such that only divisor volumes are bigger than $1$ in string units. However, empirically divisor volumes are not very sensitive to angular position in the K\"ahler cone at large $h^{1,1}$ \cite{Demirtas:2021gsq}, so we will assume that the tip of the stretched K\"ahler cone is a representative point in the moduli space. }

Another condition that we impose on the moduli arises from considering possible corrections to the tree-level K\"ahler potential \eqref{KahlerpotentialEq}. As discussed in Appendix \ref{Kahlerappendix}, ensuring that conceivable corrections to the K\"ahler potential are subleading amounts to ensuring that cross-terms in \eqref{Vfull} between single instanton contributions and the flux superpotential dominate over cross-terms between multiple instantons.  We therefore expect the effective field theory to be under computational control when, in addition to \eqref{eq:vol1}, we have
\begin{align}\label{condition}
|\qalpha \cdot \tauvec| \gg \left| \qalphaT \boldsymbol{K}^{-1} \qalphaprime \right| e^{-2 \pi \qalphaprime \cdot \tauvec} \,, ~~~ \forall \alpha \neq \alpha' \,.
\end{align}

In summary, our method to pick the moduli in a given geometry is as follows: first, we find the tip of the stretched K\"ahler cone. Then, we perform an overall dilatation of the Calabi-Yau until \eqref{condition} is imposed. 
In general, this will create a steep hierarchy \cite{Demirtas:2018akl,Demirtas:2021gsq} between the instanton scales. This hierarchy allows us to truncate the full potential \eqref{Vfull}, retaining only the $P \geq N$ terms that are sufficiently large that they could significantly affect the potential (see \S\ref{truncationsec} for details). In this way we obtain an axion potential of the form \eqref{VP}.  To ensure we investigate the regime close to the edge of perturbative control where the largest number of terms contribute, we impose \eqref{condition} as a simple inequality.

\section{Finding all minima} \label{numericalmin}
Finding critical points of functions of many variables is often intractable both analytically and numerically. In our case the equations are transcendental, and numerical techniques such as gradient descent scale  poorly with dimension.  Adding to the challenge numerically are the extremely large hierarchies in our potential.  Some components of the gradient are  exponentially larger than others at typical points in field space, and keeping track of this requires extreme numerical precision.\footnote{For clarity we will refer to this as the ``hierarchy problem".}  We surmount these difficulties by utilizing the large hierarchies to truncate the potential from infinitely many terms down to $P = \mathcal{O}(N)$.  These terms turn out to have a sparse structure, in the sense that most of the axion directions only get significant contributions from a single instanton.  This allows us to solve for the locations of critical points analytically in most directions.  The remaining directions must be extremized numerically, but the dimension of this new problem is greatly reduced.  We overcome the hierarchy problem by rescaling these directions appropriately.

\subsection{Truncating the potential} \label{truncationsec}
The first step is to identify the largest term in \eqref{Vfull} (that is, the term with the largest amplitude) and relabel it with index $i=1$. We then consider the second largest term. If its charge vector is linearly independent of the $i=1$ charge vector, we label it $i=2$. In this case the two corresponding charges $\widetilde{\boldsymbol{q}}^{(1)}$ and $\widetilde{\boldsymbol{q}}^{(2)}$ form the first two rows of a matrix we label $\Qtilde$. When the procedure is complete, $\Qtilde$ will be a full-rank $N \times N$ matrix containing the terms that make the largest contributions to all the directions in the axion field space.

If on the other hand the charge vector corresponding to the second largest term is linearly dependent on (i.e.~parallel to) $\widetilde{\boldsymbol{q}}^{(1)}$, we keep it only if its amplitude $\Lambda^4$ satisfies $|\Lambda^4|/|\Lambda_1^4| > t$, where $t$ is a numerical threshold. If it satisfies this criterion, we assign to it an index $a = 1$ and include it as the first row $\thickbar{\boldsymbol{q}}^{(1)}$ of a $(P-N) \times N$ matrix  we call $\Qbar$.

We then proceed to the next largest terms, including them in $\Qtilde$ if their charge vector is linearly independent of the rows already filled in, and, if not, in $\Qbar$ if their coefficient is at least a fraction $t$ of \emph{any} of the coefficients corresponding to the rows in $\Qtilde$ that they have non-vanishing overlap with. The procedure comes to an end when the $N$ rows of $\Qtilde$ are filled in and the coefficients of the remaining instantons are too small to pass the threshold.

Regarding the threshold condition: it is motivated by a rigorous statement that can be made about the amount of minima in $N = 1, P = 2$ theories (Appendix \ref{countingsec}). This result suggests that when the axion charges are sparse and $\mathcal{O}(1)$ (as they are in our compactifications), and the ratio of instanton scales is small, no ``qualitatively new" minima (or more generally, critical points) are created by adding subleading corrections. As we discussed in the introduction it is possible for subleading corrections to enlarge the fundamental domain, creating (new) approximate copies of pre-existing critical points in the process as we discussed in \S\ref{axfund}. This possible effect is taken into account by the factor involving the ratio of volumes in Eq.~\eqref{Nvaceq}. \\
  
\noindent To summarize: one first identifies the largest term in \eqref{Vfull} and makes it the first row of a matrix $\Qtilde$. After, one proceeds recursively, discarding the $I$th smallest contribution to \eqref{Vfull} when its charge $\boldsymbol{q}^{(I)}$ is linearly dependent on the charges already present in $\Qtilde$ and its amplitude $\Lambda_I^4$ is small enough compared to the amplitudes of the charges on which it linearly depends, i.e.~$|\Lambda_I^4|/|\Lambdatilde_i^4| < t$ for all charges $\widetilde{\boldsymbol{q}}^{(i)}$ already present in $\Qtilde$ for which $\boldsymbol{q}^{(I)} \cdot \widetilde{\boldsymbol{q}}^{(i)} \neq 0$. Otherwise, we include the contribution in the truncation, adding it to $\Qtilde$ or $\Qbar$ depending on whether the new charge is linearly independent, or dependent, respectively, of the charges already present in $\Qtilde$.

The result of this procedure is a truncated potential \eqref{VP},
\begin{align}\label{min0}
	V_P(\thetavec) &= \sum_{I=1}^P \Lambda_I^4 \,\cos \left( 2 \pi \left( \mathbfcal{Q}\thetavec \right)^I + \delta_I \right) \\
	&= \sum_{i=1}^N \Lambdatilde_i^4 \, \cos \left( 2\pi (\Qtilde \thetavec)^i + \widetilde{\delta}_i \right) + \sum_{a=1}^{P-N} \Lambdabar_a^4 \, \cos \left( 2\pi ( \Qbar \thetavec)^a + \thickbar{\delta}_a \right) \,. \label{VPbis}
\end{align}
Notice that the terms are ordered as\footnote{Where often, $\geq$ is $\ggg$ due to the exponential dependence of the coefficients on the cycle volumes in string units.}
 \begin{align}\label{min0bis}
	|\Lambdatilde_1^4| &\geq |\Lambdatilde_2^4| \geq \cdots \geq |\Lambdatilde_N^4| \,, \\
 |\Lambdabar_1^4| &\geq |\Lambdabar_2^4| \geq \cdots \geq |\Lambdabar_{P-N}^4| \,,
\end{align}
but that there is no general ordering between the $\Lambdatilde_i^4$ and the $\Lambdabar_a^4$.

\subsection{Finding the critical points} \label{criticalsec}
As we did in \S\ref{axfund}, to find the critical points of the truncated potential \eqref{VPbis} it is convenient to change coordinates to
\begin{equation}\label{coordtrans}
    \thetatildevec = \Qtilde \thetavec + \deltatilde/2 \pi \,.
\end{equation}
(This is a good change of coordinates because $\Qtilde$ is full rank, but it is not in general volume-preserving or orthogonal.) In terms of these coordinates the truncated axion potential takes the form
\begin{equation} \label{Vthetatilde}
	V_P(\thetatildevec) = \sum_{i=1}^N \Lambdatilde_i^4 \, \cos \left( 2\pi \widetilde{\theta}^i \right) + \sum_{a=1}^{P-N} \Lambdabar_a^4 \, \cos \left( 2\pi \alphavec^{(a)} \cdot \thetatildevec + \delta_a \right) \,,
\end{equation}
where $\alphavec^{(a)} = (\Qtilde^{-1})^\top \qbar^{(a)}$ and we have redefined $\deltavec = \deltabar - \Qbar \deltatilde$.  The phases $\delta_a $ cannot be removed by a field redefinition and are physical.  We do not have a way to compute them from first principles; therefore, in our numerical analysis we choose them randomly with a uniform distribution.

In the compactifications we consider it turns out the charge vectors $\alphavec^{(a)}$ are sparse and $P-N$ is not very large.  As a result, only a few axion fields $\widetilde \theta^i$ appear in the second sum in \eqref{Vthetatilde}. The majority of the axions, which appear only in the first term, can be dealt with trivially -- a fact made manifest by the coordinate transformation  \eqref{coordtrans} -- leaving a much lower-dimensional problem that can be solved numerically. In this way we were able to identify all extrema even for potentials in $\sim 500$-dimensional field spaces.

To illustrate the algorithm for finding extrema, consider the case $P-N = 1$, where the effective charge matrix in \eqref{Vthetatilde} takes the form
\begin{equation} \label{qsimple}
	\begin{pmatrix}
\mathbbold{1}_N \\
\alphavec^{(1) \top} = \left( \alpha^{(1)}_{1}, \dots, \alpha^{(1)}_{N} \right)
\end{pmatrix} \,.
\end{equation}
As we mentioned above, $M \lesssim \mathcal{O}(N)$ of the $\alpha_i^{(1)}$ vanish in the cases we studied. Considering \eqref{Vthetatilde}, this implies that a fundamental domain of $V_P$ in the $\thetatildevec$-coordinates is of the form $\mathcal{F}(V_P) = [0,1)^M \times \mathcal{F}(V_{P-M})$. Now, critical points satisfy the $N$ equations 
\begin{align} \label{rootone}
     \sin (2 \pi \widetilde{\theta}^i) +
    \frac{\Lambdabar_1^4}{\Lambdatilde_i^4} \alpha^{(1)}_{i} \sin \left( 2 \pi \alphavec^{(1)} \cdot \thetatildevec + \delta_1 \right) = 0 \,, ~~~ \forall i \in \{1, \cdots, N \} \,.
\end{align}
The equations for which $\alpha^{(1)}_{i} = 0$ have distinct solutions $\widetilde{\theta}^i = 0$ (contributing a negative Hessian eigenvalue) or $\widetilde{\theta}^i = 1/2$ (contributing a positive Hessian eigenvalue). The remaining $N-M$ equations for the remaining $N-M$ axions must be solved numerically. To do this it is important to know $\mathcal{F}(V_{P-M})$ so that one knows in which region the distinct extrema lie. This region may be computed in general via the method described in Appendix \ref{Smithappendix} and is of the form\footnote{If $\boldsymbol{\alpha}^{(1)}$ contains integers only, which turns out to occur often in our ensemble of potentials, we know $\mathcal{F}(V_{P-M}) = [0,1)^{N-M}$ is a fundamental domain for the remaining axions.}
\begin{equation} \label{Vsimplest3}
    \mathcal{F}(V_{P-M}) = \left\{ \sum_{i=1}^{N-M} p_i \boldsymbol{\beta}^{(i)} ~ \big| ~ p_i \in [0,1) \right\} \,,
\end{equation}
where the $\boldsymbol{\beta}^{(i)}$ are a basis for the shift symmetries of the $N-M$ axions in \eqref{Vthetatilde} that are not separable. Our strategy is then to solve the non-trivial equations in \eqref{rootone} by starting from $\sim 10^5$  points defined by drawing the coefficients $p_i$ in \eqref{Vsimplest3} independently from the uniform distribution on $[0,1)$. If the numerical solver finds a solution outside $\mathcal{F}(V_{P-M})$, we shift it back into $\mathcal{F}(V_{P-M})$ with the appropriate integer linear combination of the $\boldsymbol{\beta}^{(i)}$.

In fact we can optimize the root-finding procedure further still by making use of the large hierarchies in the scales $\Lambda^4$: when the ratio of scales in \eqref{rootone} is less than the threshold $t$ defined in \S\ref{truncationsec} we should be able to ignore the second term when locating extrema numerically.

In the general case $P-N \geq 1$ the critical point equations read 
\begin{equation}\label{pmN}
 \sin (2\pi \widetilde{\theta}^i) +
    \sum_{a=1}^{P-N} \frac{\Lambdabar_a^4}{\Lambdatilde_i^4} \alpha^{(a)}_i \sin \left( 2 \pi \alphavec^{(a)} \cdot \thetatildevec + \delta_a \right) = 0 \,, ~~~ \forall i \in \{ 1, \cdots, N \} \,.
\end{equation}
As above we can separate out the trivial directions, determine a fundamental domain for the remaining terms, further neglect small terms and solve the remaining transcendental equations numerically. We used a standard Python root finder algorithm for this purpose.

To summarize, the sparseness of the effective charges $\alphavec^{(a)}$ in \eqref{Vthetatilde} and the large hierarchies in the magnitudes of the axion potential terms  permit us to greatly reduce the dimensionality of the system of equations that must be solved numerically, allowing us to identify all critical points.

\subsection{Identifying the minima} \label{minimasec}
Having found all the unique critical points of the truncated potential $V_P$, we may pick out the minima by identifying those for which the Hessian is positive definite. This is complicated numerically by the large hierarchies in the scales $\Lambdatilde_i^{4}$.  This difficulty can be overcome simply by rescaling each axion $\widetilde \theta^i$  by $|\Lambdatilde_i^{4}|^{-1/2}$, which changes the Hessian from $H$ to $\hat{H}$ as
\begin{align} \label{hessscaled}
    \hat{H}_{ij} &= \frac{H_{ij}}{\sqrt{|\Lambdatilde_i^4|}\sqrt{|\Lambdatilde_j^4|}} \,.
\end{align}
This rescaling endows the matrix $\hat{H}$ with the following properties: it is symmetric, all the diagonal terms are $\mathcal{O}(1)$, all the non-diagonal terms are $\lesssim \mathcal{O}(1)$, and it preserves number of positive and negative eigenvalues of the true Hessian $H$. These properties make it easy to find the spectrum numerically and determine if a given critical point is a minimum.

After this last step we will have found the number of distinct minima of $V_P$, $\mathcal{N}_P$. Formula \eqref{Nvaceq} then determines the total amount of distinct minima of the full potential $V$ in Eq.~\eqref{Vfull}. As described in the previous section, the fundamental domain $\mathcal{F}(V_P)$ in the $\thetatildevec$-coordinates splits as $\mathcal{F}(V_P) = [0,1)^M \times \mathcal{F}(V_{P-M})$, whereas a basis for the lattice of symmetries of $V_{P-M}$ is given by an $(N-M) \times (N-M)$ matrix $\boldsymbol{\mathcal{B}}$ with the $\boldsymbol{\beta}^{(i)}$ of Eq.~\eqref{Vsimplest3} on its columns (again, see Appendix~\ref{Smithappendix} for how to find $\boldsymbol{\mathcal{B}}$ in general), often simply equal to $\mathbbold{1}_{N-M}$ in our ensemble of potentials. Thus in the $\thetatildevec$-coordinates, $\text{vol} \, \mathcal{F}(V_P) = \left| \det \boldsymbol{\mathcal{B}} \right|$. Finally, if $\mathcal{F}(V) = [0,1)^N$ in the original $\thetavec$-coordinates we used to write \eqref{Vfull}, $\text{vol} \, \mathcal{F}(V) = | \det \widetilde{\mathbfcal{Q}} |$ in the $\thetatildevec$-coordinates. Thus
\begin{equation}
    \Nvac = \left| \frac{\det \widetilde{\mathbfcal{Q}}}{\det \boldsymbol{\mathcal{B}}} \right| \Nred \,.
\end{equation}

Naively one might expect very rapid growth in the number of minima at large $N$.  For instance, if the charge matrix is chosen randomly the number of minima scales super-exponentially with $N$, even if  $P-N = \mathcal{O}(1)$ \cite{Bachlechner:2017hsj}. An exception to this occurs if the charge matrix is very sparse, which is the case here.

The conditions \eqref{condition} become more restrictive at larger $N$. This is due to the fact that with more axion directions, to ensure all of the corrections remain under perturbative control generally requires a larger CY volume. As a result the hierarchies among energy scales are enhanced at larger $N$ and we indeed have $P-N = \mathcal{O}(1)$.  However, the charge matrix is very sparse, as it includes most or all of the $N \times N$ identity matrix among its leading terms \eqref{primetoric}.  The combination of $P-N = \mathcal{O}(1)$ and a sparse charge matrix account for the relative paucity of minima.

\newpage
\section{Results}
\begin{figure}[H]
    \centering    
\includegraphics[width = 0.93\textwidth]{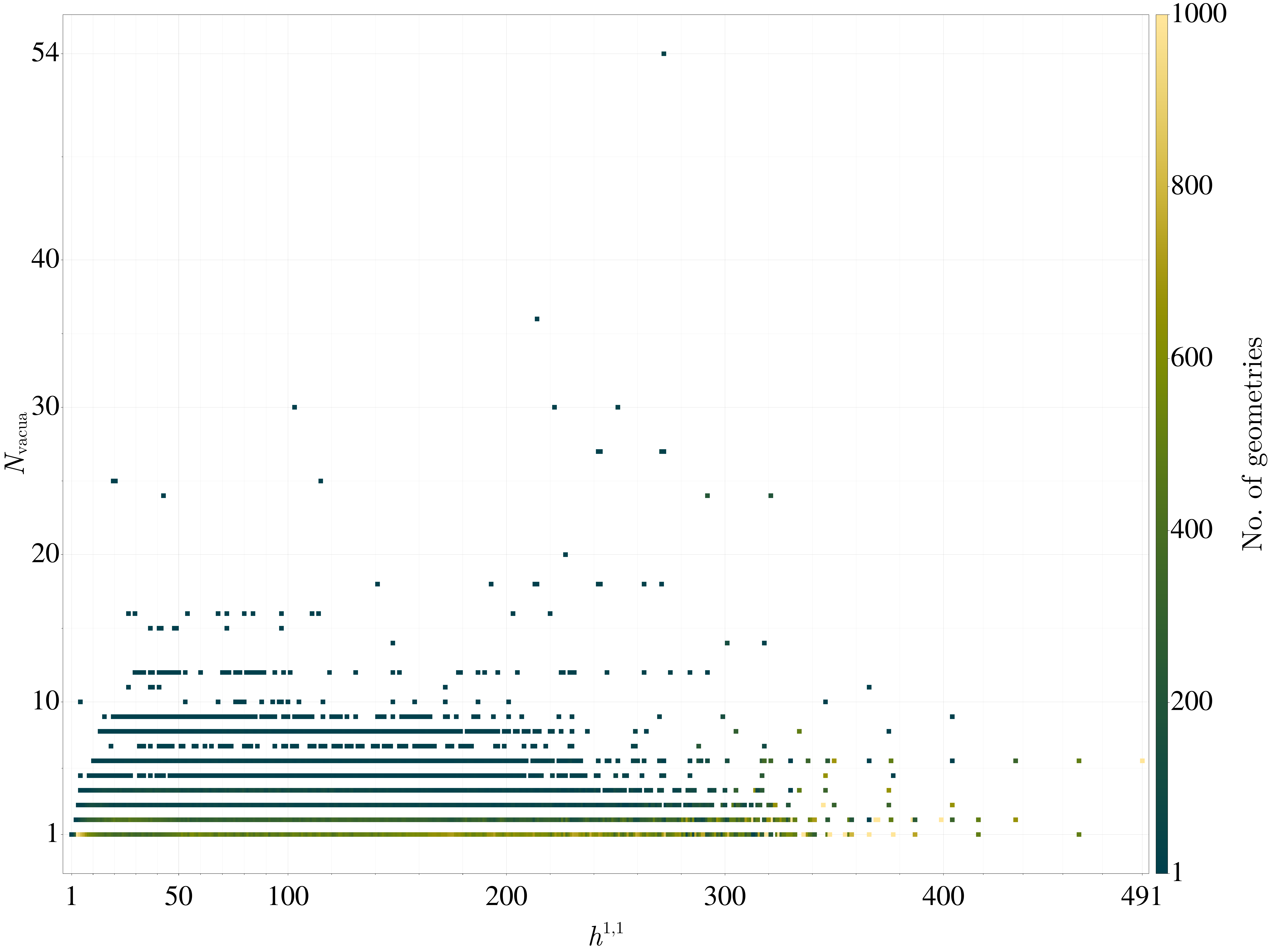}
    \caption{Distribution of numbers of distinct vacua in  $\approx 400,000$  geometries versus axion number $N = h^{1,1}$.  We sampled 1000 geometries at each $N \geq 4$ (and all geometries for smaller $N$) and computed the number of distinct minima of the axion potential for each.  The majority of geometries had only a single minimum; the maximum number of minima was 54 and the mean was 1.9. }
\label{fig:minimafig}
\end{figure}

To study the statistics of axion minima in the Kreuzer-Skarke axiverse, we employ the \texttt{CYTools} package \cite{Demirtas:2022hqf} to generate a large database of axion potentials from CY compactifications.  We use the~\texttt{random\_triangulations\_fast} routine in the \texttt{Polytope} class of \texttt{CYTools} to sample 1000 distinct geometries for each $h^{1,1} \geq 4$ (and the maximum available for smaller $h^{1,1}$) from 1 to 491.\footnote{This sampling technique produces geometries that arise from triangulations that are close together at large $h^{1,1}$, which may account for the reduced scatter in the number of minima as $h^{1,1}$ approaches 491.} In Figure \ref{fig:minimafig} we show the number of distinct axion minima for the geometries in this dataset.  Our results show that none of these compactifications have more than 54 distinct axion minima, and the great majority have either a single minimum or only a few.  The mean and median are 1.89 and 1 respectively.  This relative paucity of minima arises for the reasons discussed previously. 

It is instructive to compare these results with the numbers of minima for axion potentials where the entries of the charge matrix are drawn randomly.  For instance, if the entries in the charge matrix are $-1$ or 1 with equal probability and the matrix is taken to the square (which is appropriate given the large hierarchies present in terms in the axion potential).  The number of minima in the unit cube scales like the determinant of this random matrix, which is $ \mathcal{O}(\sqrt{N!})$, since the determinant can be written as a sum of $N!$ terms each of which is $\pm 1$ and that are approximately independent.  This is vastly larger than that of charge matrices derived from string theory.  For instance, a typical determinant for such a random matrix with $N=150$ is $\sim 10^{131}$ (see Table \ref{demo-table}).

\subsection{Numerical precision}
Our algorithm employs two approximations to estimate the  number of minima of the potential \eqref{Vfull}:
\begin{itemize}
    \item we neglect terms smaller than the threshold $t$ (\S \ref{truncationsec}),
    \item we numerically minimize the resulting truncated potential \eqref{VP}.
\end{itemize}
We believe that at least in the great majority of cases, our results for the number of minima are exact.  Typical axion potentials are difficult to minimize by gradient descent even at small $N$ due to the enormous hierarchies.  However, we tested our algorithm on artificial examples where the potential terms are all comparable in magnitude, and  brute force minimization by gradient descent with random starting points finds the same set of minima our algorithm identifies. 
  We also verified in a set of examples that changing the threshold $t$ over three orders of magnitude does not affect our results.  Lastly, we checked that including more random initial points in our numerical minimization routine does not result in finding any additional minima, and that each unique minimum is found multiple times.

\begin{table}[!h]
\begin{center}
\begin{tabular}{|c || c c c c c  |} 
 \hline
 $h^{1,1}$ & 4 & 50 & 100 & 200 & 491 \\
 \hline
 $\langle \Nvac \rangle$ & $1$ & $1.92$ & $1.95$ & $1.58$ & $6$ \\ 
 \hline
 $\sqrt{N!}$ & $10^{0.69}$ & $10^{32}$ & $10^{79}$ & $10^{187}$ & $10^{555}$ \\
 \hline
\end{tabular}
\caption{\label{demo-table}Comparison of $\langle \Nvac \rangle$, the mean number of vacua in the ensemble of string theory compactifications that we consider, to $\sqrt{N!}$, the approximate average number in an axion theory with random $\mathcal{O}(1)$ charges, as a function of the number of axion fields $N = h^{1,1}$.}
\end{center}
\end{table}

\section{Conclusions}
Our results show that individual Calabi-Yau flux compactifications -- at least those derived from the Kreuzer-Skarke database and under perturbative control -- do not give rise to very rich axion potentials with many minima.  This by itself does not mean that the landscape paradigm for solving the cosmological constant problem cannot succeed.  First, it is possible that in other compactifications, such as those   not under perturbative control or with different topology, the axion potential is richer.  Second, with $h^{1,1} + 4$ dominant terms in the charge matrix, $h^{1,1}$ of which form an identity matrix, it is very difficult to have a large number of minima. However it is possible that additional divisors involving larger charges (such as ``authochthonous" divisors) could contribute as well.  Third, even for fixed Calabi-Yau topology, different choices of flux give rise to different axion data, for instance by altering $W_0$ and changing the number of fermion zero modes.  We have not analyzed the question of how many minima might arise from varying the flux (see \cite{Denef:2004ze}). Fourth, other dynamical fields besides axions might give rise to local minima.  Fifth, there might be special points in moduli space where certain features enhance the complexity of the potential.  For instance, if many divisor volumes are equal or very close, such as in a KKLT compactification \cite{Kachru:2003aw}, there will be more terms contributing and this potentially could  give rise to more minima.  Lastly, one should really view the entire collection of compactifications with differing data as ``the landscape".  Within the set of compactifications on toric hypersurface Calabi-Yau manifolds there are enormous numbers of distinct geometries \cite{Demirtas:2018akl}.  Even if each individual geometry allows for only a few distinct axion minima, in this entire set there are a very large number of  minima.

The physics of the early universe may depend strongly on whether  in fact transitions between minima with different topology for the compactification manifold and/or different flux numbers are necessary to populate a sufficiently large landscape, versus transitions in the axion sector alone. For instance, in \cite{Bachlechner:2019vcb,Bachlechner:2018gew} it was shown that axion potentials with random parameters can produce a very large landscape that not only solves the cosmological constant problem but can explain several other macroscopic features of our universe, including inflation and dark matter.   It would be interesting to revisit this analysis and study transitions between topologies or flux numbers instead of transitions in axion potentials for fixed geometry.  

On the other hand, the sparseness of the charge matrix and the steep hierarchies in the axion potential (for a fixed geometry) is what allows the Peccei-Quinn mechanism to work well in string theory \cite{Demirtas:2021gsq}.  It also dictates axion-photon couplings, axion masses and decay constants, and affects the abundance of dark matter \cite{Demirtas:2018akl,Demirtas:2021gsq, Mehta:2020kwu,Mehta:2021pwf}.  An interesting question is whether, despite the relatively small number of minima, inflation could still take place in these axion potentials.  We intend to study this question in the future.

\paragraph{Acknowledgements:} We would like to thank M. Casiulis, M. Dentler, S. Martiniani, L. McAllister, J. Moritz, J. Roberts and P. Suryadevara for useful discussions.  The work of MK and JLM is supported by NSF grants PHY-1820814 and PHY-2112839. The work of NG is supported in part by a grant from the Simons Foundation (602883,CV), the DellaPietra Foundation, and by the NSF grant PHY-2013858, as well as by NSF grant PHY-1719877. MK and NG would like to thank the Carg\`ese School where this work was initiated. 

\appendix

\section{K\"ahler potential corrections} \label{Kahlerappendix}
The axion scalar potentials studied in this work were computed under the assumption that the superpotential and K\"ahler potential are given by \eqref{superpotentialEq} and \eqref{KahlerpotentialEq}, respectively.  As explained in \S\ref{stringsec}, due to the holomorphicity of the superpotential, corrections to the superpotential can come only from instantons supported on inherited or autochthonous effective divisors. The K\"ahler potential, on the other hand, is a real function and can in principle receive perturbative corrections, as well as non-perturbative corrections from instantons wrapping non-effective divisors. The true K\"ahler potential is then
\begin{align}
    \mathcal{K} = \mathcal{K}_{\text{tree}} + \mathcal{K}_{\text{pert.}} + \mathcal{K}_{\text{non-pert.}}
\end{align}

Let us first turn our attention to contributions to $\mathcal{K}_{\text{non-pert}}$. Such corrections can arise if and only if they have the requisite 4 fermion zero modes to contribute to the K\"ahler potential. Therefore, instantons on holomorphic divisor classes cannot alone contribute to $\mathcal{K}_{\text{non-pert}}$, but instantons supported on linear combinations of holomorphic and anti-holomorphic divisor classes can, i.e. instantons on non-effective divisors. The actions of these instantons are given by the volume of the divisors that they wrap, but these divisors are not calibrated by the K\"ahler form, and there is no practical way of computing the  volume-minimizing representative, $\Sigma_{\text{min}}$ \cite{a5b3c95f-5023-31ba-970e-30087498906c, bams/1183550129}.

One may compare the volumes of non-holomorphic divisors to the so-called ``piecewise-calibrated" representatives in the same divisor class. Given a non-holomorphic divisor class $[D]$, a piecewise-calibrated representative $\Sigma_U$  of $[D]$ is a representative in that class that is the union of a holomorphic and an anti-holomorphic divisor. The volume of $\Sigma_U$ is given by the sum of the constituent divisors:
\begin{align}
    \text{vol}(\Sigma_U) = \text{vol}(\Sigma_{\text{hol}}) + \text{vol}(\Sigma_{\text{anti-hol}}) \,.
\end{align}
We can think of the volume-minimizing representative of $[D]$ as a ``fusing" of the holomorphic and anti-holomorphic constituents, and compare the true volume of $\Sigma_{\text{min}}$ to the piecewise-calibrated volume. In \cite{Demirtas:2019lfi}, the recombination fraction was defined as
\begin{align}
  \mathfrak{r} \equiv \frac{\text{vol}(\Sigma_U)-\text{vol}(\Sigma_{\text{min}})}{\text{vol}(\Sigma_{\text{min}})}  \,.
\end{align}
A large value of $\mathfrak{r}$ means that the volume of the minimum-volume representative is much smaller than the piecewise-calibrated representative. There is no direct evidence for large recombination in the literature \cite{MicallefWolfson}, although arguments have been made using the Weak Gravity Conjecture \cite{Demirtas:2019lfi} that in certain regimes of moduli space there may be recombination.

The possibility does remain that there may be small non-holomorphic cycles that significantly alter the scalar potential arising in the effective theories considered in this work. Such a situation would invalidate not only the results presented in this work, but in all other analyses of the scalar potential. Such a discovery would be very interesting, but lies beyond the scope of this work.

We would also like to ensure that the non-perturbative corrections are subleading to the perturbative ones. It is not known how to compute perturbative corrections to the K\"ahler potential in general (although see \cite{Kim:2023eut, Kim:2023sfs} for progress in this direction), but we can attempt to estimate the regime in moduli space where they are still dominant to their non-perturbative counterparts. The goal of this endeavor is to arrive at the condition \eqref{condition} that we impose in our scans as proxy for control of the K\"ahler potential.\footnote{We thank J. Moritz for help and discussions on this topic.}

To get a handle on this, we will assume that any term of the form of one that appears in the scalar potential may also appear in the K\"ahler potential. We make this assumption because for instantons to appear in $\mathcal{K}$ they must have the requisite 4 fermion zero modes to soak up the superspace integral. Cross-terms in the scalar potential then have just the right form to correct $\mathcal{K}$, as they involve 2 zero modes from the superpotential contribution and 2 from the anti-superpotential contribution \cite{Buican:2008qe}. We will also assume, as before, that all divisor volumes are bigger than or equal to $1$, to ensure control of the $\alpha'$ expansion.

We will now show that when \eqref{condition} is not satisfied, there is a possibility that non-perturbative terms to $\mathcal{K}$ dominate over the perturbative ones. To see this let us assume that \eqref{condition} is not satisfied, i.e.
\begin{align}
    |\qalpha \cdot \tauvec| < \left| \qalphaT \boldsymbol{K}^{-1} \qalphaprime \right| e^{-2 \pi \qalphaprime \cdot \tauvec} \,,
\end{align}
for some $\alpha, \alpha'$. Multiplying both sides by $e^{-2\pi \boldsymbol{q}^{(\alpha')} \cdot \tauvec}$ we obtain
\begin{align}
      |\qalpha \cdot \tauvec|e^{-2\pi \boldsymbol{q}^{(\alpha')} \cdot \tauvec}  < \left| \qalphaT \boldsymbol{K}^{-1} \qalphaprime \right| e^{-2 \pi (\qalphaprime +\boldsymbol{q}^{(\alpha)} )\cdot \tauvec} \,,
\end{align}
and we notice that this inequality is actually comparing the first term in \eqref{Vfull} with the second term. Therefore, one may reasonably expect that these types of terms could appear in the K\"ahler potential.

Rearranging and using the fact that we are assuming that all divisors have volumes bigger than or equal to $1$ in string units, we find
\begin{align}
    \frac{1} {|\qalpha \cdot \tauvec| \left| \qalphaT \boldsymbol{K}^{-1} \qalphaprime \right|} < e^{-2\pi \boldsymbol{q}^{(\alpha')} \cdot \tauvec}
\end{align}
The LHS of this inequality takes the form of a perturbative correction to $\mathcal{K}$, while the RHS takes the form of a non-perturbative correction. We therefore conclude that imposing \eqref{condition} is a reasonable assumption to be in the regime of moduli space where the K\"ahler potential is under control. 

\section{\texorpdfstring{$\text{vol} \, \mathcal{F}(V_P) ~|~ \text{vol} \, \mathcal{F}(V)$}{volPtruncdividesvolPfull}} \label{integervacuaappendix}
In this appendix we show that the ratio of volumes appearing in Eq.~\eqref{Nvaceq} for the total amount of distinct axion minima is always an integer. More generally, the fundamental domain of any full rank truncation of the (full) potential always fits an integer amount of times into the fundamental domain of the full potential. The essential observation is that the edges of the fundamental domain of the full potential are given by \textit{integer} linear combinations of the edges of the fundamental domain of the truncated potential.

Let $\boldsymbol{A} \in \mathbb{Q}^{M \times N}$ be a collection of $M$ rows of $\mathbfcal{Q}$, which define a truncation of the full potential by including only the terms with those charges in the sum \eqref{VP} (here we have written the full potential $V$ of Eq.~\eqref{Vfull} in the form of Eq.~\eqref{VP}, with $P \leq \infty$). Additionally we ask for this truncation to be full rank, more specifically $\text{rank} ~ \boldsymbol{A} = N$, so that surely $N \leq M \leq P \leq \infty$. Then, pick out an edge of the full fundamental domain, which is specified by an integer $P$-vector $\boldsymbol{p}_{(i)}$. This point lies on the intersection of the hyperplane $\{ \mathbfcal{Q} \thetavec \}$ with the integer $P$-lattice $\mathbb{Z}^P$. In the $\thetavec$-coordinates this edge is given by a $\thetavec_{(i)}$ which satisfies
\begin{equation} \label{intersectioneq2}
	\mathbfcal{Q} \thetavec_{(i)} = \boldsymbol{p}_{(i)} \,.
\end{equation}
The solution to this equation can be expressed as
\begin{equation} \label{thetapthetat}
	\thetavec_{(i)} = \sum_{j=1}^N n_{(i),j} \thetatildevec_{(j)} \,,
\end{equation}
where the $\{ \thetatildevec_{(j)} \, | \, j = 1, \cdots,N \}$ are a basis for the intersection lattice $\{ \boldsymbol{A} \thetavec \, | \, \thetavec \in \mathbb{R}^N \} \cap \mathbb{Z}^M$ (so they form the edges of the fundamental domain of the truncated potential) and the $n_{(i),j}$ are \textit{integers}. This is because $M$ of the $P$ equations in \eqref{intersectioneq2} read $\boldsymbol{A} \thetavec_{(i)} = \boldsymbol{m}_{(i)}$, where $\boldsymbol{m}_{(i)}$ is the integer $M$-vector consisting of the appropriate $M$ entries of $\boldsymbol{p}_{(i)}$. Thus $\thetavec_{(i)}$ lies on the intersection lattice mentioned above and can therefore be expressed as an integer linear combination of its basis vectors as in \eqref{thetapthetat}. Define the matrix $\boldsymbol{N}$ by $N_{ij} = n_{(i),j}$. Then Eq.~\eqref{thetapthetat} can be written as
\begin{equation}
	\boldsymbol{\Theta} = \widetilde{\boldsymbol{\Theta}} \boldsymbol{N}^\top \,,
\end{equation}
where $\boldsymbol{\Theta}, \widetilde{\boldsymbol{\Theta}}$ are the $N \times N$ matrices with the $\thetavec_{(i)}$ respectively $\thetatildevec_{(i)}$ on their columns. That is, they are the bases for the lattice of symmetries of the full potential and the truncated potential, respectively. The ratio $\text{vol} \, \mathcal{F}(V)/ \text{vol} \, \mathcal{F}(V_P)$ we are interested in is $|\det \boldsymbol{\Theta} / \det \widetilde{\boldsymbol{\Theta}} \,|$, equal to $| \det \boldsymbol{N} \,|$, an integer.

\section{Lattice bases via the Smith normal decomposition} \label{Smithappendix}
Let $\mathbfcal{Q} \in \mathbb{Q}^{P \times N}$ define a rank $N$ truncation of the axion potential as in Eq.~\eqref{VP}. Our calculation in \S\ref{axfund} of the amount of axion minima involves finding a basis for the lattice defined by the intersection of the hyperplane $\{ \mathbfcal{Q} \thetavec \, | \, \thetavec \in \mathbb{R}^N \}$ with $\mathbb{Z}^P$. So we'd like to find all integer $P$-vectors $\boldsymbol{p}$ for which there exists a $\thetavec \in \mathbb{R}^N$ such that
\begin{equation} \label{intersectioneq}
	\mathbfcal{Q} \thetavec = \boldsymbol{p} \,.
\end{equation}
For any nonzero \textit{integer} $P \times N$ matrix $\mathbfcal{I}$ there exists a ``Smith normal form", i.e. a decomposition \cite{SNF}
\begin{equation} \label{SmithNormal}
	\boldsymbol{U} \mathbfcal{I} \boldsymbol{V} = \boldsymbol{D} \,,
\end{equation}
where $\boldsymbol{U},\boldsymbol{V}$ are unimodular $P \times P$ and $N \times N$ matrices respectively, and $\boldsymbol{D}$ is a diagonal $P \times N$ matrix with non-negative integers $d_i | d_{i+1}$ on the diagonal. The amount of nonzero $d_i$ is equal to the rank of $\mathbfcal{I}$. This will be $N$ in our case, since we will consider
\begin{equation}
    \mathbfcal{I} = \mathfrak{n} \mathbfcal{Q} \,,
\end{equation}
where $\mathfrak{n}$ is any integer that renders $\mathbfcal{I}$ integer-valued. Calling $\thetavec' \equiv \boldsymbol{V}^{-1} \thetavec/\mathfrak{n}$, \eqref{intersectioneq} is rewritten as
\begin{equation} \label{Ueq}
	\boldsymbol{U}^{-1} \begin{pmatrix}
d_1 \theta_1' \\
d_2 \theta_2' \\
\vdots \\
d_N \theta_N' \\
0 \\
\vdots \\
0
\end{pmatrix} = \boldsymbol{p} \,.
\end{equation}
If we multiply both sides of this equation by $\boldsymbol{U}$, we see that the $d_i \theta_i'$ must be integers. So \eqref{Ueq} tells us that the $\boldsymbol{p}$ that can be reached are some integer linear combinations of the first $N$ columns in $\boldsymbol{U}^{-1}$; call these $\boldsymbol{U}^{-1}_{(N)}$. These columns are all filled with integers: since $\boldsymbol{U}$ is unimodular, so too is $\boldsymbol{U}^{-1}$ in particular it is integer-valued. Thus, \textit{all} integer linear combinations of the columns in $\boldsymbol{U}^{-1}_{(N)}$ can be reached (since the $d_i \theta_i'$ can be arbitrary integers), and so these columns form a basis for the intersection lattice. In the $\thetavec$-coordinates a set of basis vectors, placed on the columns of a matrix $\mathbfcal{B}$, is obtained from these by multiplying $\boldsymbol{U}^{-1}_{(N)}$ by $(\mathbfcal{Q}^\top \mathbfcal{Q})^{-1} \mathbfcal{Q}^\top$ (e.g. \cite{Bachlechner:2017hsj} -- this can be seen by solving for $\thetavec$ in \eqref{intersectioneq}): $\mathbfcal{B} = (\mathbfcal{Q}^\top \mathbfcal{Q})^{-1} \mathbfcal{Q}^\top \boldsymbol{U}^{-1}_{(N)}$. The volume $\mathcal{F}(V_P)$ (or $\mathcal{F}(V)$, if one chooses not to truncate; $P = \infty$) of the unit cell defined by this basis, appearing in our formula \eqref{Nvaceq} for the total amount of distinct axion minima, is then finally (e.g. \cite{latticeintro})
\begin{equation}
    \text{vol} \, \mathcal{F}(V) = \left| \det \mathbfcal{B} \right| = \left| \frac{\det \mathbfcal{Q}^\top \boldsymbol{U}^{-1}_{(N)}}{\det \mathbfcal{Q}^\top \mathbfcal{Q}} \right| \,.
\end{equation}
When $P=N$, $\boldsymbol{U}^{-1}_{(N)} = \mathbbold{1}_N$ and this formula is seen to reduce to Eq.~\eqref{volFV} of the main text.

\section{Exact result in \texorpdfstring{$N=1, P = 2$}{Nis1Pis2} potentials} \label{countingsec}
In this appendix we prove the following theorem about the amount of distinct minima in $N = 1, P = 2$ axion theories. It concretizes our intuition that small amplitude perturbations only lift the degeneracy of pre-existing minima (i.e.~they do not create ``new" minima), provided their frequency is not too large.
\begin{remark}
The single-axion potential
\begin{equation} \label{Vtheorem}
	V(\theta) = \Lambda_1^4 \cos(2 \pi q_1 \theta + \delta_1) + \Lambda_2^4 \cos(2 \pi q_2 \theta + \delta_2) \,,
\end{equation}
where $\Lambda_1^4, \Lambda_2^4 \in \mathbb{R}_0$, $\delta_1, \delta_2 \in \mathbb{R}$, $q_1, q_2 \in \mathbb{Q}_0$ and
\begin{equation}
	|\eta| + |\widetilde{\eta}| < 1
\end{equation}
where
\begin{equation}
	\eta = \frac{q_2 \Lambda_2^4}{q_1 \Lambda_1^4} \,, ~~~ \widetilde{\eta} = \frac{q_2}{q_1} \eta \,,
\end{equation}
has $A$ distinct minima, where $A/B = |q_1/q_2|$ is in irreducible form.
\end{remark}

\begin{proof}
	We may take $q_1, q_2 > 0$ and $\delta_1 = 0$ without loss of generality, so that
\begin{equation}
	V(\theta) = \Lambda_1^4 \cos(2 \pi q_1 \theta) + \Lambda_2^4 \cos(2 \pi q_2 \theta + \delta) \,.
\end{equation}
As a first step we note the period of $V$, namely $A/q_1 = B/q_2$, where $A$ and $B$ are smallest possible integers. Thus $q_1/q_2 = A/B \in \mathbb{Q}$, where this last fraction is in irreducible form. Our task is to find the distinct minima of $V$, i.e. those inside one fundamental domain, say $\theta \in [0,A/q_1)$. We will assume $\Lambda_2^4 > 0 > \Lambda_1^4$ -- the proof in the other cases is analogous.
	
Minima of $V$ occur at those $\theta$ for which
\begin{align}
	\sin x &= -\eta \sin y \,, \label{Vprime} \\
	\cos x &> -\widetilde{\eta} \cos y \,, \label{Vdoubleprime}
\end{align}
where $x = 2 \pi q_1 \theta$, $y = 2 \pi q_2 \theta + \delta$. From \eqref{Vprime} it follows that minima must be located in symmetric intervals around $x = n \pi$, $n \in \mathbb{Z}$, with diameter $\pi |\eta|$. This is because for $x \in [0,\pi/2]$, $\sin x \geq 2x/\pi$, and so if $\pi/2 \geq x > \pi |\eta|/2$, $\sin x > |\eta|$ but this is incompatible with \eqref{Vprime} (by symmetry of $\sin$ the same argument applies around all other $x = n \pi$, both to the right and to the left of these points). Similarly, from \eqref{Vdoubleprime} it follows that minima must be located at a distance at most $\pi/2 + \pi |\widetilde{\eta}|/2$ from $x = 2 \pi m$, $m \in \mathbb{Z}$. The possible minima near odd multiples of $\pi$ are then ruled out because $|\eta| + |\widetilde{\eta}| < 1$. Thus we are left with possible minima around $x = 2 \pi m$ in intervals of diameter $\pi |\eta|$. In these intervals we are interested in the condition \eqref{Vprime}, i.e. in the intersection(s) of the functions $\sin \alpha$ and $-\eta \sin \left( 2 \pi m B/A + \delta + q_2 \alpha / q_1 \right) = -\eta \sin \left( q_2 \alpha / q_1 + \delta' \right)$ for $|\alpha| < \pi |\eta|/2$, and all $m = 0,1,\cdots,A-1$, where we have written $x = 2\pi m + \alpha$. There is certainly at least one intersection between these functions on that interval: the first function starts on the left from a negative value $\sin(-\pi |\eta|/2) < -|\eta|$ and ends on the right at a positive value $\sin(\pi |\eta|/2) > |\eta|$. The second function starts left at a value larger than $-|\eta|$ and ends right at a value smaller than $|\eta|$. So the functions must cross at least once -- this follows from applying the intermediate value theorem to the difference of the functions.

It remains to be shown that there is exactly one crossing, which also satisfies \eqref{Vdoubleprime}. Regarding the latter, notice that
\begin{equation}
	\cos x \geq 1-|\eta| > |\widetilde{\eta}| \geq - \widetilde{\eta} \, \cos y \,,
\end{equation}
where the first inequality follows from the concavity of $\cos$ near its maximum. So \eqref{Vdoubleprime} is always satisfied at the crossing points. Regarding the former, redefining variables as we did before, we need to show that $\sin \alpha$ and $-\eta \sin \left( 2 \pi m B/A + \delta + q_2 \alpha / q_1 \right) = -\eta \sin \left( q_2 \alpha / q_1 + \delta' \right)$ have a single crossing point in the interval $|\alpha| < \pi |\eta|/2$, for any $m = 0,\cdots,A-1$. It would suffice if
\begin{equation}
	\sin \alpha = -\eta \sin \left( \frac{q_2}{q_1} \alpha + \delta' \right)
\end{equation}
had a single solution in that interval, for any $\delta' \in [0,2\pi)$. We proceed by contradiction: let's assume these functions had at least one other intersection. First, we notice that the right-most intersection must happen at an $\alpha \geq 0$. Indeed, either the left-most intersection already happens at $\alpha \geq 0$, and then so too does the right-most, or the left-most intersection happens at an $\alpha < 0$. In that case, immediately after the first crossing, we would have\footnote{We thank M. Mirbabayi for help with this part of the proof.}
\begin{equation} \label{greatercondition}
	\sin \alpha > -\eta \sin \left( \frac{q_2}{q_1} \alpha + \delta' \right) \,.
\end{equation}
Squaring, we obtain $\sin^2 \alpha < \eta^2 \sin^2 \beta$ where $\beta = q_2 \alpha/q_1 + \delta'$ (the inequality is reversed because $\alpha < 0$), from which it follows that $\cos^2 \alpha > 1 - \eta^2 + \eta^2 \cos^2 \beta$, from which in turn it follows that $\cos^2 \alpha > \widetilde{\eta}^2 \cos^2 \beta$. Therefore $\cos \alpha > |\widetilde{\eta} \cos \beta| \geq -\widetilde{\eta} \cos \beta$. This means that the derivative of the function on the LHS of \eqref{greatercondition} is greater than the derivative of the function on the RHS of \eqref{greatercondition} for $\alpha < 0$, from which we conclude that the inequality \eqref{greatercondition} will be maintained for all $\alpha < 0$. Therefore, the right-most intersection could only occur at $\alpha \geq 0$. Finally, just beyond this right-most intersection, we would have
\begin{align}
	-\eta \sin \beta \geq \sin \alpha \,, \\
	-\widetilde{\eta} \cos \beta \geq \cos \alpha \,.
\end{align}
Squaring both sides, the inequalities maintain their direction since $\alpha \geq 0$, and, adding them up yields $\eta^2 \sin^2 \beta + \widetilde{\eta}^2 \cos^2 \beta \geq 1$. This is a contradiction since $\eta^2 \sin^2 \beta + \widetilde{\eta}^2 \cos^2 \beta \leq \eta^2 + \widetilde{\eta}^2 < |\eta| + |\widetilde{\eta}| < 1$.
\end{proof}

\newpage
\vfill
\bibliographystyle{klebphys2}
\bibliography{refs}
\end{document}